\documentclass[preprint,superscriptaddress,nofootinbib]{revtex4}
\usepackage[dvips]{graphicx}
\usepackage{amsmath,amssymb}

\def\lromn#1{\uppercase\expandafter{\romannumeral#1}}

\begin{document}

  \preprint{UT-HET 074}


  \title{Higgs inflation in a radiative seesaw model} 
  \author{Shinya Kanemura}
  \email{kanemu@sci.u-toyama.ac.jp}
  \affiliation{
  Department of Physics,
  University of Toyama, Toyama 930-8555, Japan
  }
  \author{Toshinori Matsui} 
  \email{matsui@jodo.sci.u-toyama.ac.jp}
  \affiliation{
  Department of Physics,
  University of Toyama, Toyama 930-8555, Japan
  }
  \author{Takehiro Nabeshima}
  \email{nabe@jodo.sci.u-toyama.ac.jp}
  \affiliation{
  Department of Physics,
  University of Toyama, Toyama 930-8555, Japan
  }
  \begin{abstract}
  We investigate a simple model to explain inflation, neutrino masses and dark matter simultaneously. 
  This is based on the so-called radiative seesaw model proposed by Ma 
  in order to explain neutrino masses and dark matter by introducing 
  a $Z_2$-odd isospin doublet scalar field and $Z_2$-odd right-handed neutrinos. 
  We study the possibility that the Higgs boson as well as neutral components of 
  the $Z_2$-odd scalar doublet field can satisfy conditions 
  from slow-roll inflation and vacuum stability up to the inflation scale. 
  We find that a part of  parameter regions where these scalar fields can play a role of an inflaton 
  is compatible with the current data from neutrino experiments and 
  those of the dark matter abundance as well as the direct search results. 
  A phenomenological consequence of this scenario results in a specific mass spectrum of scalar bosons, 
  which can be tested at the LHC, the International Linear Collider and the Compact Linear Collider. 
  \end{abstract} 
\maketitle

\section{Introduction}


The new particle with the mass of 126~GeV which has been found at the LHC~\cite{atlas, cms} 
is showing various properties that the Higgs boson must have. 
It is likely that the particle is the Higgs boson. 
If this is the case, the Standard Model (SM) of elementary particles is confirmed 
its correctness not only in the gauge interaction sector but also in the sector of electroweak symmetry breaking. 
By the discovery of the Higgs boson, all the particle contents in the SM are completed. 
This means that we are standing on the new stage to search for new physics beyond the SM. 
There are several empirical reasons why we consider the new physics. 
Phenomena such as neutrino oscillation~\cite{solar, atom, acc-disapp, acc-app, short-reac, long-reac}, 
existence of dark matter~\cite{WMAP} and 
baryon asymmetry of the Universe~\cite{WMAP, Sakharov:1967dj, Cyburt:2008kw} cannot be explained in the SM. 
Cosmic inflation at the very early era of the Universe~\cite{inf}, which is a promising candidate to solve 
cosmological problems such as the horizon problem and the flatness problem, 
also requires the additional scalar boson, the inflaton. 

The determination of the Higgs boson mass at the LHC opens 
the door to directly explore the physics at very high scales. 
Assuming the SM with one Higgs doublet, the vacuum stability argument indicates that 
the model can be well defined only below the energy scale where 
the running coupling of the Higgs self-coupling becomes zero. 
For the Higgs boson mass to be 126~GeV with the top quark mass to be 173.1~GeV 
and the coupling for the strong force to be $\alpha_s =$ 0.1184, 
the critical energy scale is estimated to be around $10^{10}$~GeV by the NNLO calculation, 
although the uncertainty due to the values of the top quark mass and $\alpha_s$ is not small~\cite{lambda_run}. 
The vacuum seems to be metastable when we assume that the model holds up to the Planck scale. 
This kind of analysis gives a strong constraint on the scenario of the Higgs inflation~\cite{Hinf} where 
the Higgs boson works as an inflaton, because the inflation occurs at the energy scale where 
the vacuum stability is not guaranteed in the SM. 
Recently, a viable model for the Higgs inflation has been proposed, 
in which the Higgs sector is extended including an additional scalar doublet field~\cite{2Hinf}. 

In order to generate tiny masses of neutrinos, various kinds of models have been proposed. 
The simplest scenario is so called the seesaw mechanism, 
where the tiny neutrino masses are generated at the tree level by introducing very heavy particles, such as 
right-handed neutrinos~\cite{type-I}, a complex triplet scalar field~\cite{type-II}, 
or a complex triplet fermion field~\cite{type-III}. 
The radiative seesaw scenario is an alternative way to explain tiny neutrino masses, where 
they are radiatively induced at the one loop level or at the three loop level 
by introducing $Z_2$-odd scalar fields and $Z_2$-odd right-handed neutrinos~\cite{KNT, Ma, AKS}. 
An interesting characteristic feature in these radiative seesaw models is that 
dark matter candidates automatically enter into the model because of the $Z_2$ parity. 

In this Letter, 
we discuss a simple model to explain inflation, neutrino masses and dark matter simultaneously, 
which is based on the simplest radiative seesaw model~\cite{Ma}. 
Both the Higgs boson and neutral components of 
the $Z_2$-odd scalar doublet can satisfy conditions 
for slow-roll inflation~\cite{slow-roll} and vacuum stability up to the inflation scale. 
We find that a part of the parameter region where these scalar fields can play a role of the inflaton 
is compatible with the current data from neutrino experiments and 
those of the dark matter abundance as well as the direct search results~\cite{XENON100}. 
A phenomenological consequence of scenario results in a specific mass spectrum of scalar fields, 
which can be tested at the LHC, the International Linear Collider (ILC)~\cite{ILC} and
the Compact Linear Collider (CLIC)~\cite{CLIC}. 

\section{Lagrangian}
\begin{table}[t]
\begin{center}
\begin{tabular}
{|p{25mm}|@{\vrule width 1.8pt\ }p{15mm}|p{15mm}|p{15mm}|p{15mm}|p{15mm}|p{15mm}|p{15mm}|p{15mm}|}
   \hline
     &$Q_L$&$u_R$&$d_R$&$L_L$&$\ell_R$&$\Phi_1$&$\Phi_2$&$\nu_R^{}$
    \\ \noalign{\hrule height 1.8pt}
    SU(3)$_{\rm C}$&{\bf 3}&{\bf 3}&{\bf 3}&{\bf 1}&{\bf 1}&{\bf 1}&{\bf 1}&{\bf 1}
    \\ \hline
    SU(2)$_{\rm I}$&{\bf 2}&{\bf 1}&{\bf 1}&{\bf 2}&{\bf 1}&{\bf 2}&{\bf 2}&{\bf 1}
    \\ \hline
    U(1)$_{\rm Y}$&$\frac{1}{6}$&$\frac{2}{3}$&$-\frac{1}{3}$&$-\frac{1}{2}$&$-$1&$\frac{1}{2}$&$\frac{1}{2}$&0
    \\ \hline
    $Z_2$&1&1&1&1&1&1&$-$1&$-$1
    \\ \hline
\end{tabular}
\caption{Particle contents and their quantum charges.}
\label{table:particle}
\end{center}
\end{table}
We consider the model, which is invariant under the unbroken discrete $Z_2$ symmetry, 
with the $Z_2$-odd scalar doublet field $\Phi_2$ 
and right-handed neutrino $\nu_R^{}$ to the SM with the SM Higgs doublet field $\Phi_1$~\cite{Ma}. 
Quantum charges of particles in the model are shown in Table~\ref{table:particle}. 
Dirac Yukawa couplings of neutrinos are forbidden by the $Z_2$ symmetry. 
The Yukawa interaction for leptons is given by 
\begin{eqnarray}
{\cal L}_{Yukawa} &=& Y_\ell \overline{L_L}\Phi_1\ell_R+Y_\nu\overline{L_L}\Phi_2^c\nu_R+h.c.,
\end{eqnarray}
where the superscript $c$ denotes the charge conjugation. 
The scalar potential is given by~\cite{2Hinf} 
\begin{eqnarray}
V &=&\frac{M_P^2 R}{2}+(\xi_1|\Phi_1|^2 +\xi_2 |\Phi_2|^2)R 
\nonumber\\
  &&+
\mu_1^2 |\Phi_1|^2 + \mu_2^2 |\Phi_2|^2 + \frac{1}{2} \lambda_1 |\Phi_1|^4 + \frac{1}{2} \lambda_2 |\Phi_2|^4 + \lambda_3|\Phi_1|^2|\Phi_2|^2 
\nonumber\\
  &&+ 
\lambda_4 (\Phi_1^{\dagger} \Phi_2) (\Phi_2^{\dagger} \Phi_1) 
+ [\frac{1}{2}\lambda_5((\Phi_1^{\dagger} \Phi_2)^2+h.c.)], 
\label{eq:potential}
\end{eqnarray}
where $M_P$ is the Planck scale ($M_P\simeq 10^{19}$~GeV), and $R$ is the Ricci scalar.

We assume that $\mu_1^2 < $0 and $\mu_2^2 >$ 0. 
$\Phi_1$ obtains the vacuum expectation value (VEV) $v$ ($=\sqrt{-2\mu_1^2/\lambda_1}\simeq 246{\rm GeV}$), 
while $\Phi_2$ cannot get the VEV because of the unbroken $Z_2$ symmetry. 
The lightest $Z_2$-odd particle is stabilized by the $Z_2$ parity, and 
it can act as the dark matter as long as it is electrically neutral. 
The quartic coupling constants should satisfy the following constraints 
on the unbounded-from-below conditions at the tree level; 
\begin{eqnarray}
\lambda_1>0,\ \ \lambda_2>0,\ \ \lambda_3+\lambda_4+\lambda_5+\sqrt{\lambda_1 \lambda_2}>0. 
 \label{eq:vs}
\end{eqnarray}
Three Nambu-Goldstone bosons in the Higgs doublet field $\Phi_1$ are absorbed by the $Z$ and $W$ bosons
by the Higgs mechanism. 

Mass eigenstates of the scalar bosons are 
the SM-like $Z_2$-even Higgs scalar boson ($h$), 
the $Z_2$-odd CP-even scalar boson ($H$), 
the $Z_2$-odd CP-odd scalar boson ($A$) and 
$Z_2$-odd charged scalar bosons ($H^\pm$). 
Masses of these scalar bosons are given by~\cite{Ma}
\begin{eqnarray}
m_h^2
&=&
\lambda_1 v^2, 
\nonumber \\
m_H^2
&=&
\mu_2^2 +\frac{1}{2}(\lambda_3+\lambda_4+\lambda_5) v^2,
\nonumber \\
m_A^2
&=&
\mu_2^2 +\frac{1}{2}(\lambda_3+\lambda_4-\lambda_5) v^2,
\nonumber \\
m_{H^{\pm}}^2
&=&
\mu_2^2 +\frac{1}{2}\lambda_3 v^2.
 \label{eq:hmass}
\end{eqnarray}
\section{Constraint on the model from inflation and dark matter}
\subsection{Inflation}
We consider the Higgs inflation scenario~\cite{Hinf,2Hinf,other_Hinf} in our model
defined in the previous section. 
The scalar potential is given in the Einstein frame by 
\begin{eqnarray}
V_E &\simeq&\frac{\lambda_1+\lambda_2r^4+2(\lambda_3+\lambda_4)r^2+2\lambda_5r^2\cos(2\theta)}{8(\xi_2r^2+\xi_1)^2}
\left(1-e^{-2\phi/\sqrt{6}}\right)^2, 
 \label{eq:potentialE}
\end{eqnarray}
where $\phi, r$ and $\theta$ are defined as
\begin{equation}
\Phi_1= \frac{1}{\sqrt{2}}
\begin{pmatrix}
 0  \\
 h_1
\end{pmatrix}
, \ \ \ 
\Phi_2= \frac{1}{\sqrt{2}}
\begin{pmatrix}
 0  \\
 h_2e^{i\theta}
\end{pmatrix},
\nonumber
\end{equation}
\begin{eqnarray}
\phi= \sqrt{\frac{3}{2}}\ln(1+\frac{\xi_1h_1^2}{M_P^2}+\frac{\xi_2h_2^2}{M_P^2}),\ \ \ 
r=\frac{h_2}{h_1},
 \label{eq:fieldE}
\end{eqnarray}
with taking a large field limit $\xi_1h_1^2/M_P^2+\xi_2h_2^2/M_P^2 \gg 1$.

For stabilizing $r$ as a finite value, 
we need to impose following conditions~\cite{2Hinf}; 
\begin{eqnarray}
\lambda_2\xi_1-(\lambda_3+\lambda_4)\xi_2&>&0,
\nonumber\\
\lambda_1\xi_2-(\lambda_3+\lambda_4)\xi_1&>&0,
\nonumber\\
\lambda_1\lambda_2-(\lambda_3+\lambda_4)^2&>&0.
 \label{eq:vs2}
\end{eqnarray}
Parameters in the scalar potential should satisfy 
the constraint from the power spectrum~\cite{WMAP, 2Hinf};
\begin{eqnarray}
\xi_2 \sqrt{\frac{2(\lambda_1+a^2\lambda_2-2a(\lambda_3+\lambda_4))}{\lambda_1\lambda_2-(\lambda_3+\lambda_4)^2}}
&\simeq& 5\times 10^{4},
 \label{eq:xi}
\end{eqnarray}
\begin{eqnarray}
\frac{\lambda_5}{\xi_2} 
\frac{a\lambda_2 - (\lambda_3+\lambda_4)}{\lambda_1+a^2\lambda_2-2a(\lambda_3+\lambda_4)}
&\lesssim& 4\times 10^{-12},
 \label{eq:l5}
\end{eqnarray}
where $a$ is given as $a\equiv\xi_1/\xi_2$.
When the scalar potential satisfies the conditions in Eqs. (\ref{eq:vs2})-(\ref{eq:l5}), 
the model could realize the inflation. 
\subsection{Dark Matter}
We assume that the CP-odd boson $A$ is the lightest $Z_2$ odd particle. 
(By changing the sign of the coupling constant $\lambda_5$, 
the similar discussion can be applied with the CP-even boson $H$ to be the lightest.) 
When $\lambda_5$ is very small such as ${\cal O}(10^{-7})$, 
$A$ is difficult to act as the dark matter 
because the scattering process $AN\to HN$ opens, where $N$ is a nucleon. 
The cross section is too large to be consistent 
with the current direct search results for dark matter~\cite{direct_Z, Kashiwase:2012xd, LopezHonorez:2006gr}. 
In Ref.~\cite{2Hinf}, the authors claim that both the Higgs boson and 
$Z_2$-odd neutral scalar bosons can work as the inflatons 
when the dark matter ($H$ or $A$) has the mass of 600~GeV if $\lambda_5 \lesssim 10^{-7}$. 
However, as recently discussed in Ref.~\cite{Kashiwase:2012xd}, 
the bound from direct search results are getting stronger, and 
such a dark matter is not allowed anymore in this model without a fine tuning 
among the scalar self-coupling constants. 
We here take $\lambda_5\simeq 10^{-6}$ and 
\begin{eqnarray}
a\lambda_2 - (\lambda_3+\lambda_4)\simeq 10^{-1}
\label{eq:FT}
\end{eqnarray}
at the inflation scale. 
With this choice, the process $AN\to HN$ can be avoided kinematically. 
Still masses of $A$ and $H$ are almost the same value. 
The coannihilation process $AH\to XX$ via the $Z$ boson is important 
to explain the abundance of the dark matter where X is a  particle in the SM, 
because the pair annihilation process $AA\to XX$ via the $h$ boson is suppressed due to 
the constraint from the inflation. 
The cross section of $AH\to XX$ depends only on the mass of the dark matter. 
Therefore, the mass of the dark matter $A$ is constrained from the abundance of the dark matter as 
\begin{eqnarray}
128~{\rm GeV}\leq m_A\leq138~{\rm GeV}, 
 \label{eq:WMAP}
\end{eqnarray}
where we have used the nine years WMAP data~\cite{WMAP}
.
The scattering process $AN\to AN$ then comes mainly from the diagram of the SM-like Higgs boson mediation. 
The cross section is given by~\cite{DM_direct, LopezHonorez:2006gr} 
\begin{eqnarray}
\sigma(AN\to AN)\simeq\frac{\lambda_{hAA}^2}{4m_h^4}\frac{m_N^2}{\pi(m_A+m_N)^2}f_N^2, 
 \label{eq:AN}
\end{eqnarray}
where
$\lambda_{hAA}
\equiv
\lambda_3+\lambda_4-\lambda_5$, $f_N
\equiv
\sum_{q}m_Nf_{Tq}-\frac{2}{9}m_Nf_{TG}$ and 
$m_N$ is the mass of nucleon, where 
$f_{Tu}+f_{Td}=0.056$, $f_{Ts}=0$~\cite{lattice} and $f_{TG}=0.944$~\cite{trace-anomaly}.
The mass $m_A$ should be approximately a half of $m_h$~\cite{inertDM} 
in order for the dark matter to be consistent with 
the abundance from the WMAP experiment~\cite{WMAP} 
and the upper bound on the scattering cross section for $AN\to AN$ 
from the XENON100 experiment~\cite{XENON100}. 
The coupling constant $\lambda_{hAA}$ should satisfy 
\begin{eqnarray}
\lambda_{hAA}\lesssim 0.3,
\label{eq:hAA}
\end{eqnarray}
at the low energy scale for consistency to satisfy the data from the XENON100 experiment. 
\subsection{Tiny Neutrino Masses}
\begin{figure}[t]
 \begin{center}
  \scalebox{0.2}{\includegraphics{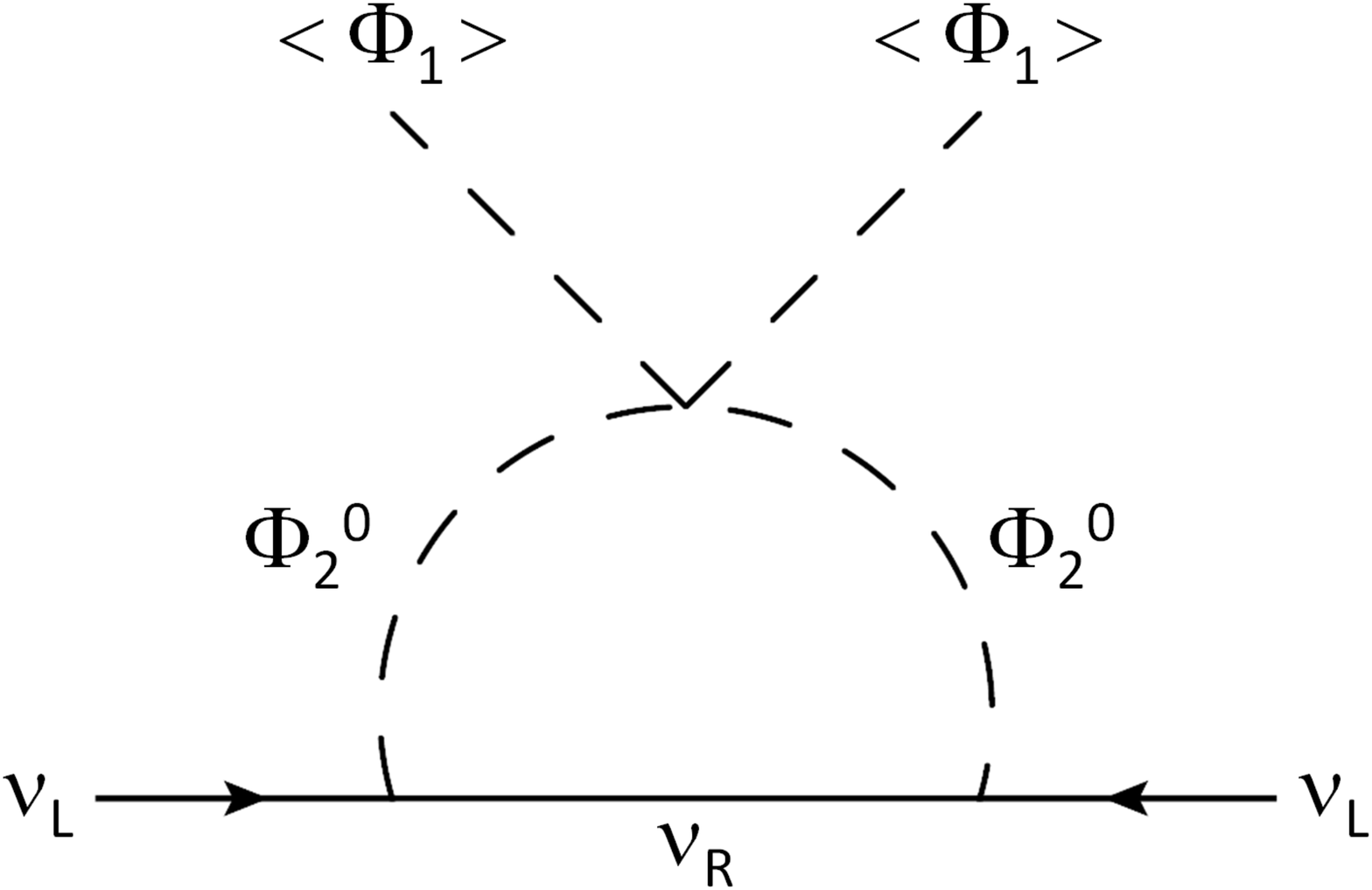}}
  \caption{The Feynman diagram for tiny neutrino masses.}
  \label{fig:neutrinomass}
 \end{center}
\end{figure}

In this model, tiny neutrino masses are generated by 
the one loop diagram in Fig.~\ref{fig:neutrinomass}~\cite{Ma}. 
The neutrino mass $(m_{\nu})_{ij}$ are given by 
\begin{eqnarray}
(m_{\nu})_{ij}
=
\sum_{k}
\frac{(Y_\nu)_i^k(Y_\nu)_j^kM_R^k}{16\pi^2}
\left[\frac{m_H^2}{m_H^2-\left(M_R^k\right)^2}\ln\frac{m_H^2}{\left(M_R^k\right)^2}
-\frac{m_A^2}{m_A^2-\left(M_R^k\right)^2}\ln\frac{m_A^2}{\left(M_R^k\right)^2}\right],
 \label{eq:nmass}
\end{eqnarray}
where $M_R^k$ is the Majorana mass of $\nu_R^k$ ($k$=1-3). 
The flavor structure of $(m_\nu)_{ij}$ is given by $(Y_\nu)_i^k(Y_\nu)_j^k/M_R^k$. 
The neutrino mixing matrix is explained by neutrino Yukawa coupling constants $(Y_\nu)_i^k$. 
The magnitude of tiny neutrino masses can be explained 
when $(Y_\nu)_i^k(Y_\nu)_j^k/M_R^k\simeq {\cal O}(10^{-7})$~GeV$^{-1}$
because $\lambda_5$ and masses of scalar bosons, $m_H^{}$ and $m_A^{}$, are constrained from the conditions of 
inflation and dark matter. 
Our model then can be consistent with current experimental data 
for neutrinos~\cite{solar, atom, acc-disapp, acc-app, short-reac, long-reac}. 
For example, 
when $M_R^k$ is ${\cal O}(1)$~TeV, $(Y_\nu)_i^k$ is ${\cal O}(10^{-2})$.

\subsection{Running of Scalar Coupling Constants}
In the SM, the energy scale cannot reach to the inflation scale because 
the quartic coupling constant of the Higgs boson is inconsistent with 
the unbounded-from-below condition at $10^{10}$~GeV scale 
when $m_t$ = 173.1~GeV and $\alpha_s$ = 0.1184~\cite{lambda_run}. 
On the other hand, if we consider extended Higgs sectors such as the two Higgs doublet model, 
the vacuum stability condition on the quartic coupling constant for the SM-like Higgs boson can be relaxed 
due to the effect of the additional quartic coupling constants~\cite{extended_run}. 
Therefore, these models can be stable up to the inflation scale\footnote{
When we consider $(Y_{\nu})_i^k$ is ${\cal O}(10^{-2})$, 
the contribution from the right-handed neutrino loop is as negligible as that from $b$ quarks.} 
We calculate these coupling constants by using 
the renormalization group equations with the following beta functions~\cite{beta}; 
\begin{align}
\beta(g_s)&=\frac{-7g_s^3}{16\pi^2},\ \ \ 
\beta(g)=\frac{-3g^3}{16\pi^2},\ \ \ 
\beta(g')=\frac{7g^{'3}}{16\pi^2},\\
\beta(y_t)&=\frac{y_t}{16\pi^2}\Big[\frac{9}{2}y_t^2-8g_s^2-\frac{9}{4}g^2-\frac{17}{12}g^{'2}\Big],
\end{align}
\begin{align}
\beta(\lambda_1)&=\frac{1}{16\pi^2}\Big[12\lambda_1^2+4\lambda_3^2+2\lambda_4^2+2\lambda_5^2+4\lambda_3\lambda_4-12y_t^4+12y_t^2\lambda_1
\notag\\
&\quad\quad+\frac{9}{4}g^4+\frac{3}{2}g^2g^{'2}+\frac{3}{4}g^{'4}-3\lambda_1\big(3g^2+g^{'2}\big)\Big],\\
\beta(\lambda_2)&=\frac{1}{16\pi^2}\Big[12\lambda_2^2+4\lambda_3^2+2\lambda_4^2+2\lambda_5^2+4\lambda_3\lambda_4
+\frac{9}{4}g^4+\frac{3}{2}g^2g^{'2}+\frac{3}{4}g^{'4}-3\lambda_2\big(3g^2+g^{'2}\big)\Big],\\
\beta(\lambda_3)&=\frac{1}{16\pi^2}\Big[6\lambda_1\lambda_3+2\lambda_1\lambda_4+6\lambda_2\lambda_3+2\lambda_2\lambda_4+4\lambda_3^2+2\lambda_4^2+2\lambda_5^2\notag\\
&\quad\quad+\frac{9}{4}g^4+\frac{3}{4}g^{'4}-\frac{3}{2}g^2g^{'2}-3\lambda_3\big(3g^2+g^{'2}\big)+6\lambda_3y_t^2\Big],
\end{align}
\begin{align}
\beta(\lambda_4)&=\frac{1}{16\pi^2}\Big[2\lambda_4(\lambda_1+\lambda_2+4\lambda_3+2\lambda_4)+8\lambda_5^2+3g^2g^{'2}
-3\lambda_4\big(3g^2+g^{'2}\big)+6\lambda_4y_t^2\Big],\\
\beta(\lambda_5)&=\frac{1}{16\pi^2}\Big[2\lambda_5(\lambda_1+\lambda_2+4\lambda_3+6\lambda_4)-3\lambda_5(3g^2+g^{'2})+6\lambda_5y_t^2\Big]. 
\end{align}
\begin{figure}[t]
 \begin{center}
  \scalebox{0.6}{\includegraphics{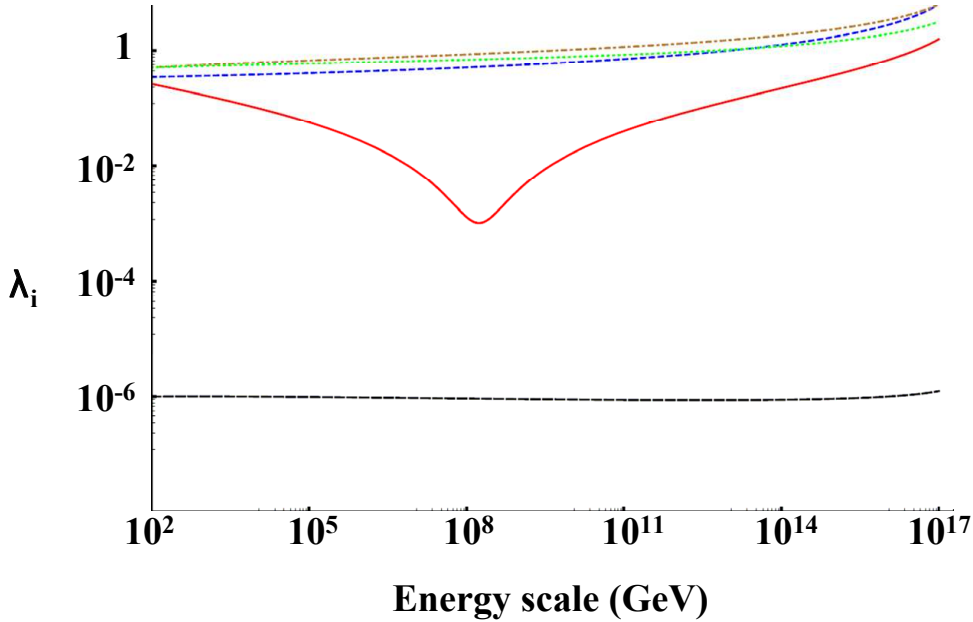}}
  \caption{Running of the scalar coupling constants.
  Red (solid), blue (dashed), brown (dot-dashed), green (dotted) and black (long-dashed) curves show 
  $\lambda_1$, $\lambda_2$, $\lambda_3$, $-\lambda_4$ and $\lambda_5$, respectively.}
  \label{fig:run}
 \end{center}
\end{figure} 
We here impose the conditions of triviality 
\begin{eqnarray}
\lambda_i \lesssim 2\pi, 
\label{eq:tri}
\end{eqnarray}
and vacuum stability (the unbounded-below-condition) up to the inflation scale. 
In Fig~\ref{fig:run}, running of the scalar coupling constants are shown 
between the electroweak scale and the inflation scale. 
The vacuum instability due to $\lambda_1$ is avoided by the effect of the Higgs self-coupling constants with 
$Z_2$-odd scalar bosons~\cite{extended_run}. 
In Table~\ref{table:lambda}, 
we show an example for the values of the scalar coupling constants 
at the scales of ${\cal O}(10^{2})$~GeV and ${\cal O}(10^{17})$~GeV, 
which satisfy the conditions of the inflation and the dark matter, 
where ${\cal O}(10^{17})$~GeV denotes the inflation scale for $\xi_1\simeq\xi_2={\cal O}(10^{4})$~\cite{Hinf, 2Hinf} 
in our model\footnote{
When $\xi_1\simeq\xi_2={\cal O}(10^{4})$, unitarity is broken at $M_P/\xi_1$~\cite{uni_break}. 
Then, we should introduce new particle at the unitarity breaking scale to save unitarity~\cite{uni_care}. 
However, we do not consider the effect of this particle on the  running of $\lambda$ coupling constants 
because this effect affect only above ${\cal O}(10^{15})$~GeV. 
The effect is expected to be smaller than the effect of the SM particles. 
}. 
\subsection{Mass Spectrum}
Let us evaluate the mass spectrum of the model under the constraint from inflation, 
the neutrino data and the dark matter data as well as the vacuum stability condition. 
In our model, there are nine parameters in the scalar sector; 
i.e., $\xi_1$, $\xi_2$, $\mu_1^2$, $\mu_2^2$, 
$\lambda_1$, $\lambda_2$, $\lambda_3$, $\lambda_4$ and $\lambda_5$. 

First of all, as the numerical inputs, we take $v = 246~$GeV and $m_h = 126$~GeV. 
Second, we use the conditions to explain the thermal fluctuation; 
i.e., 
the allowed region for the mass of the dark matter $A$ 
is determined from the constraint of the dark matter abundance from the WMAP data in Eq.~(\ref{eq:WMAP}).
We here take $m_A^{} = 130$~GeV as a reference value. 
Further numerical input comes from the perturbativity of $\lambda_2$ up to the inflation scale; 
i.e., $\lambda_2(\mu_{\rm inf}) = 2\pi$, where $\mu_{\rm inf}$ is the inflation scale $10^{17}$~GeV. 
The parameter set in Table~\ref{table:lambda} can be consistent with these numerical inputs and 
the constraints are given in Eqs.~(\ref{eq:vs}), (\ref{eq:vs2}), (\ref{eq:xi}), (\ref{eq:l5}), (\ref{eq:FT})
, (\ref{eq:hAA}) and (\ref{eq:tri}). 
The mass spectrum of the scalar bosons is determined as 
\begin{eqnarray}
m_h
&\simeq&
126~{\rm GeV}, 
\nonumber \\
m_{H^{\pm}}
&\simeq&
173~{\rm GeV},
\nonumber \\
m_H
&\simeq&
130~{\rm GeV},
\nonumber \\
m_A
&\simeq&
130~{\rm GeV},
 \label{eq:hmassD}
\end{eqnarray}
where the mass difference between $A$ and $H$ is about 500~KeV.
\begin{table}[t]
\begin{center}
\begin{tabular}{|c|c|c|c|c|c|}
\hline
&$\lambda_{1}$&$\lambda_{2}$&$\lambda_{3}$&$\lambda_{4}$&$\lambda_{5}$\\
\hline
 $10^{2}$~GeV
  &0.26 &0.35&0.51&-0.51&1.0$\times10^{-6}$\\
\hline
 $10^{17}$~GeV
  &1.6 &6.3&6.3&-3.2&1.2$\times10^{-6}$\\
\hline
\end{tabular}
\caption{An example for the parameter set which satisfies 
constraints from the inflation and the dark matter 
at the scales of ${\cal O}(10^{2})$~GeV and ${\cal O}(10^{17})$~GeV. 
}
\label{table:lambda}
\end{center}
\end{table}

The mass spectrum is not largely changed even if $m_A^{}$ is varied with in its allowed region. 
Consequently, in our scenario, the following relation for the mass is obtained; 
\begin{eqnarray}
m_{H^\pm}\simeq m_A+40~{\rm GeV}. 
 \label{eq:hmassDconst}
\end{eqnarray}
The bounds on $m_{H^\pm}$ is obtained in order to satisfy the conditions from 
Eqs.~(\ref{eq:vs}), (\ref{eq:vs2}) and (\ref{eq:tri}). 
Therefore, we can test the model by using the mass spectrum at collider experiments. 
\section{Phenomenology}
Masses of $Z_2$-odd scalar bosons have been constrained by the LEP experiment. 
In our scenario, $m_{H^\pm}^{}$ should be around 170~GeV, 
which is above the lower bound given by the LEP experiment~\cite{LEP_direct, LEP_pm}. 
From the $Z$ boson width measurement, 
$m_{H}^{} + m_{A}^{}$ should be larger than $m_Z^{}$~\cite{LEP_direct, LEP_HApair}. 
In addition, there is a bound on $H$$A$ production at the LEP. 
However, when $m_{H}^{} - m_{A}^{} < 8$~GeV, masses of neutral $Z_2$-odd scalar bosons are not
constrained by the LEP~\cite{LEP_direct, LEP_HApair}.
The contributions to the electroweak parameters~\cite{STdef} 
from additional scalar bosons loops are given by~\cite{ST1, ST2} 
\begin{eqnarray}
\Delta S &=& 
-\frac{1}{4\pi}
\left[
F'_\Delta(m_{H^\pm}^{}, m_{H^\pm}^{})-F'_\Delta(m_{H}^{}, m_{A}^{})
\right],
\\
\Delta T &=& 
-\frac{\sqrt{2}G_F}{16\pi^2\alpha_{EM}}
\left[
-F_\Delta(m_{A}^{}, m_{H^\pm}^{})-F_\Delta(m_{H}^{}, m_{H^\pm}^{})+F_\Delta(m_{H}^{}, m_{A}^{})
\right],
 \label{eq:ST}
\end{eqnarray}
where 
\begin{eqnarray}
F_\Delta(x, y) &=& F_\Delta(y, x) = 
\frac{x^2+y^2}{2}-\frac{x^2y^2}{x^2-y^2}\ln\frac{x^2}{y^2}, 
\\
F'_\Delta(x, y) &=& F'_\Delta(y, x) = 
-\frac{1}{3}
\left[
\frac{4}{3}-\frac{x^2\ln x^2-y^2\ln y^2}{x^2-y^2}-\frac{x^2+y^2}{(x^2-y^2)^2}F_\Delta(x, y)
\right]. 
 \label{eq:F}
\end{eqnarray}
In all of our parameters, it is consistent with current electroweak precision data 
with 90\% Confidence Level (C.L.)~\cite{ST2}.

The detectability of $H, A$ and $H^\pm$
at the LHC has been studied in Ref.~\cite{LHC1, LHC2, LHC3}. 
They conclude that it could be difficult to test 
$pp\to AH^+/HH^+/H^+H^-$ processes because 
cross sections of the background processes are very large. 
The process of $pp\to AH$ could be tested with about the 3$\sigma$ C.L. 
with the various benchmark points for masses for $A$ and $H$ 
However, it would be difficult to test 
$pp\to AH$ in our model. 
In our parameter set, $m_H$ and $m_A$ are about 130~GeV. 
In this case, after imposing the basic cuts~\cite{LHC1, LHC2, LHC3}, event number of $pp\to AH$ is negligibly small. 
Furthermore, the total decay width of $H$ is about $10^{-29}$~GeV. 
In this case, $H$ would pass through the detector. 
Therefore, this signal is difficult to be detected at the LHC. 

\begin{figure}[t]
 \begin{center}
  \scalebox{0.5}{\includegraphics[angle=-90]{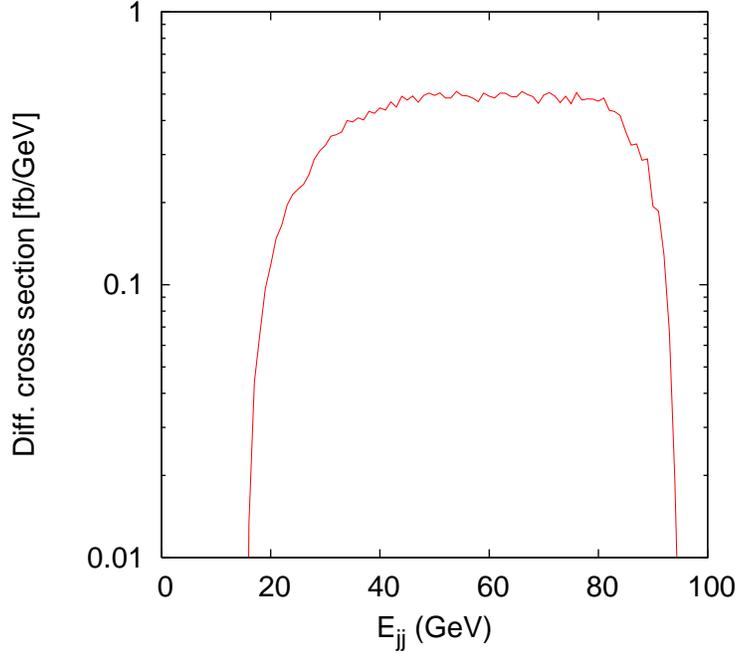}}
  \caption{The distribution of $E_{jj}$ for the cross section for 
  $e^+e^-\to H^+H^-\to W^{+(*)}W^{-(*)}AA\to jj\ell\nu AA$.}
  \label{fig:Ejj}
 \end{center}
\end{figure}

We now discuss signals of $H, A$ and $H^\pm$ at the ILC with $\sqrt{s}=500$~GeV. 
In the following, we use Calchep~2.5.6 for numerical evaluation~\cite{calc}.
First, we focus on the $H^\pm$ pair production process 
$e^+e^-\to Z^*(\gamma^*)\to H^+H^-\to W^{+(*)}W^{-(*)}AA\to jj\ell\nu AA$, 
where $j$ denotes a hadron jet~\cite{inILC}. 
The final state of this process is a charged lepton and two jets with a missing momentum. 
The energy of the two-jet system $E_{jj}$ satisfies 
the following equation because of the kinematical reason; 
\begin{eqnarray}
\frac{m_{H^{\pm}}^2-m_A^2}{\sqrt{s}+2\sqrt{s/4-m_{H^{\pm}}^2}}
<
E_{jj}
<
\frac{m_{H^{\pm}}^2-m_A^2}{\sqrt{s}-2\sqrt{s/4-m_{H^{\pm}}^2}}. 
 \label{eq:Ejj}
\end{eqnarray}
$E_{jj}$ is evaluated by using our parameter set as 
\begin{eqnarray}
15~{\rm GeV}
<
E_{jj}
<
94~{\rm GeV}.
 \label{eq:Ejj_set}
\end{eqnarray}
The distribution of $E_{jj}$ of the cross section for 
$e^+e^-\to Z^*(\gamma^*)\to H^+H^-\to W^{+(*)}W^{-(*)}AA\to jj\ell\nu AA$ 
is shown in Fig.~\ref{fig:Ejj}.
The important background processes against 
$e^+e^-\to Z^*(\gamma^*)\to H^+H^-\to W^{+(*)}W^{-(*)}AA\to jj\ell\nu AA$ are 
$e^+e^-\to W^{+}W^{-}\to jj\ell\overline{\nu}$ and 
$e^+e^-\to Z(\gamma )Z\to jj\ell\overline{\ell}$ with a missing $\overline{\ell}$ event.
In these processes, 
the missing invariant mass is zero. 
These backgrounds could be well reduced by imposing an appropriate kinematic cuts. 
We expect that $m_{H^\pm}$ and $m_A$ can be measured by using the endpoints of $E_{jj}$ at the ILC 
after the background reduction. 

Second, we focus on $HA$ production 
$e^+e^-\to Z^*\to HA\to AAZ^*\to AAjj$ 
at the ILC. 
When the mass difference between $A$ and $H$ is sizable, 
it could also be detected by using the endpoint of $E_{jj}$. 
However, 
in our mass spectrum, it is predicted that masses of $A$ and $H$ are almost degenerated. 
When we detect $H^\pm$ but we cannot detect the clue of this process at the ILC, 
it seems that masses of $A$ and $H$ are almost same value. 

Finally, we discuss prediction on the diphoton decay of the Higgs boson $h$. 
BR($h\to\gamma\gamma$) in the model, which the SM with $Z_2$-odd scalar doublet, has been studied in Ref.~\cite{hgg}. 
The deviation in our model from the SM is given by 
\begin{eqnarray}
\frac{BR(h\to\gamma\gamma)}{BR(h_{SM}\to\gamma\gamma)}
&=&
\frac{\left|N_c Q_f^2 A_{1/2}(\tau_t)+A_1(\tau_W)+\frac{\lambda_3 v^2}{2 m_{H^\pm}^2}A_0(\tau_{H^{\pm}})\right|^2}
{\left|N_c Q_f^2 A_{1/2}(\tau_t)+A_1(\tau_W)\right|^2}, 
 \label{eq:hgg}
\end{eqnarray}
where $N_c$ and $Q_f$ are the color and electromagnetic charges of the top quark, respectively. 
$A_{1/2}(x)$, $A_1(x)$ and $A_0(x)$ denote 
\begin{eqnarray}
A_{1/2}(x)
&=&
2\left\{x+(x-1)f(x)\right\}x^{-2}, 
\nonumber \\
A_{1}(x)
&=&
-\left\{2x^2+3x+3(2x -1)f(x)\right\}x^{-2}, 
\nonumber \\
A_{0}(x)
&=&
-\left\{x-f(x)\right\}x^{-2}, 
 \label{eq:A}
\end{eqnarray}
where $\tau_x$ and $f(x)$ are given by 
\begin{eqnarray}
\tau_x
&=&
\left(\frac{m_h}{2 m_x}\right)^2, 
\\
f(x)
&=&
\begin{Bmatrix}
{\rm arcsin}^2(\sqrt{x}) &x\leq 1 \\
-\frac{1}{4}\left[\ln\left(\frac{1+\sqrt{1-1/x}}{1-\sqrt{1-1/x}}\right)-i\pi\right]^2 &x\geq 1
\end{Bmatrix}. 
 \label{eq:A}
\end{eqnarray}
When we use our parameter set in Eq.~(\ref{eq:hmassD}), the ratio is calculated as 
\begin{eqnarray}
\frac{BR(h\to\gamma\gamma)}{BR(h_{SM}\to\gamma\gamma)}
&=&
0.95. 
 \label{eq:hggsol}
\end{eqnarray}
In our model, BR($h\to\gamma\gamma$) is smaller than the SM results 
due to constraints from the conditions of the inflation and the dark matter. 
These ratio is at most 10 \% because our model contains only one charged scalar field. 
\section{Discussion and Conclusion}
In this Letter, we have not explicitly discussed baryogenesis. 
It is likely not difficult to complement the mechanism for baryogenesis to our model 
via leptogenesis~\cite{Fukugita:1986hr}. 
In Ref.~\cite{Kashiwase:2012xd}, the possibility of the leptogenesis in the Ma model~\cite{Ma} 
has been studied in details under the constraint of current neutrino and dark matter data. 
By using the typical value for $\lambda_5$ in our model 
$\lambda_5\simeq 10^{-6}$), 
the scenario of baryogenesis through the leptogenesis would be difficult 
if masses of the right-handed neutrinos are about $1$~TeV. 

On the other hand, the possibility of electroweak baryogenesis would also be interesting~\cite{EWBG_org}. 
The condition of strong first order phase transition is compatible with $m_h = 126$~GeV in the framework of 
two Higgs doublet models~\cite{EWBG_2HDM} including the inert doublet model~\cite{EWBG_IDM}. 
In such a case, an important phenomenological consequence is a large deviation 
in the loop-corrected prediction on the $hhh$ coupling~\cite{EWBG_hhh}, 
by which the scenario can be tested when the $hhh$ coupling is measured 
at future colliders such as the ILC or the CLIC. 
However, in the inert doublet model including the model we have discussed in this Letter, 
an extension has to be needed in order to get additional CP violating phases, 
which are required for successful baryogenesis. 

We have studied the simple scenario to explain inflation, neutrino masses and dark matter simultaneously 
based on the radiative seesaw model with the Higgs inflation mechanism. 
We find that the parameter region where $Z_2$-odd scalar fields can play a role of the inflaton 
is compatible with the current data from neutrino experiments and 
those of the dark matter abundance as well as the direct search results. 
This scenario predicts a specific mass spectrum for the scalar fields, 
which can be measured at the LHC and the ILC with $\sqrt{s}=500$~GeV. 
Our model is a viable example for the TeV scale model for inflation (and neutrino with dark matter) 
which is testable at collider experiments.

\begin{acknowledgments}
We would like to thank Hiroshi Yokoya for the useful discussion. 
The work of S.K. was supported in part by Grant-in-Aid for Scientific Research, 
Nos. 22244031, 23104006 and 24340046. 
The work of T.N. was supported in part by 
the Japan Society for the Promotion of Science as a research fellow (DC2).
\end{acknowledgments}

\end{document}